# Rule 30: Solving the Chaos

Mayukhmali Das
*Electronics and Telecommunication*
*Jadavpur University*
Kolkata, India
mayukhmalidas322@gmail.com

Abstract: ***This paper provides an analytical solution to the Wolfram Alpha Problem 1. In this paper we discuss whether the central column of the Rule 30 structure is purely random and aperiodic.***

## 1. INTRODUCTION

Rule 30 is an elementary cellular automaton discovered by Stephen Wolfram [1]. The number 30 is significant as its binary representation actually gives us the state transition logic. Table 1 explains the logic behind the state transition. The fourth column of Table 1 when read bottom to top, gives us binary number 00011110 which is 30 in decimal. The basic Boolean logic behind rule 30 can be represented as

$$left\_cell \wedge centre\_cell\ |\ right\_cell \text{ ------- (i)}$$

The most interesting fact about Rule 30 is that its central column is hypothesized to produce completely random values and also being not periodic [2]. In this paper we will discuss the analytic solution to this hypothesis in parts and finally draw a conclusion. First, we will discuss what is giving rise to the randomness in case of Rule 30.

## 2. REASON BEHIND THE RANDOMNESS

Elementary Cellular Automata are governed by simple equations and most of them produce symmetric, systematic values. Let's consider case of Rule 150 whose Boolean expression is very much similar to Rule 30

$$left\_cell \wedge centre\_cell \wedge right\_cell \text{ -- (ii)}$$

Fig 1 shows the plot of Rule 150 for first 1024 iterations. We can see that the figure is symmetric about an axis and also has got repeating patterns throughout. This nature of Rule 150 is due to the fact that Equation (ii) has no decisive branches.

| Present state for Left cell | Present state for Centre Cell | Present state for Right Cell | Next state for Centre Cell |
|---|---|---|---|
| 0 | 0 | 0 | 0 |
| 0 | 0 | 1 | 1 |
| 0 | 1 | 0 | 1 |
| 0 | 1 | 1 | 1 |
| 1 | 0 | 0 | 1 |
| 1 | 0 | 1 | 0 |
| 1 | 1 | 0 | 0 |
| 1 | 1 | 1 | 0 |

Table 1: State Transition Logic for Center Cell

Decisive branching is the branching of a Boolean expression into another equation of different nature based on a particular condition of the expression. Equation (i) has decisive branches. We know that the two-bit truth tables of XOR and OR are both identical, except the case where both the bits equal to one. Which means equation (i) corresponding to Rule 30 can be replaced by symmetric equation (ii) of Rule 150; when both right cell and center cell are not equal to one. But when they are equal to one, the output is given by the complement of the value of the left cell. It is this transition between a symmetric state and another state that is giving rise to the randomness. Thus, in rule 30, the decisive branching is caused by the relation

$$centre\ cell\ \&\ right\_cell \text{ ------- (iii)}$$

Fig 2 shows the state machine diagram, which demonstrates the phenomenon of how decisive branching takes place in Rule 30.

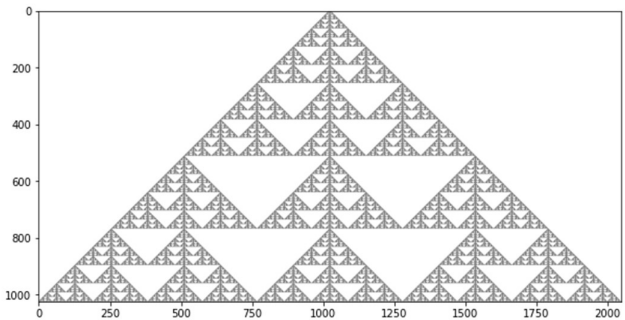

Fig 1. Rule 150 for first 1024 iteration

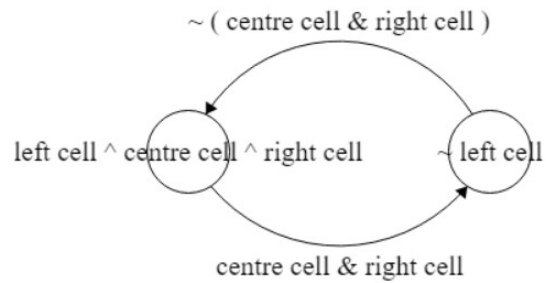

Fig 2. Decisive Branching in Rule 30

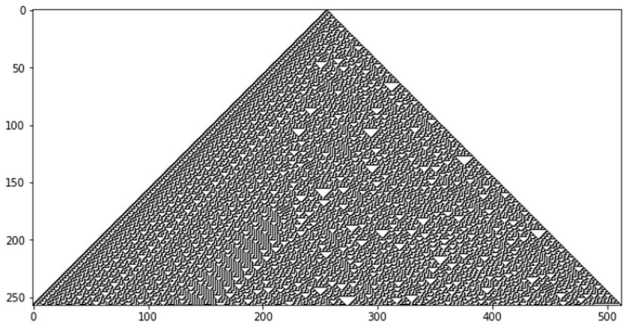

Fig 3. Rule 30 for 256 iterations

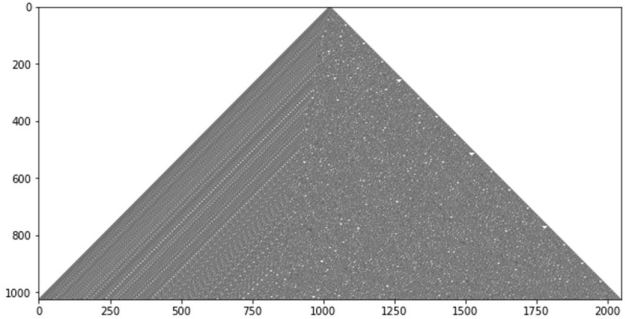

Fig 4. Rule 30 for 1024 iterations

Fig 3 and Fig 4 are the plots of Rule 30 for 256 and 1024 iterations respectively. It is evident from the figures that the left side is symmetric. This is due to the fact that decisive branching equation given in Equation (iii) involves only right and center cells. So, the left part is not involved in the randomness. We can cross-verify this by using the Boolean expression

$$left\_cell \mid centre\_cell \wedge right\_cell \quad \text{------- (iv)}$$

This expression just exchanges the OR and XOR signs of Rule 30. From Fig 5 it is evident that the symmetric part is towards the right side as the decisive branching is now being governed by

$$centre\ cell\ \&\ left\_cell \quad \text{------- (v)}$$

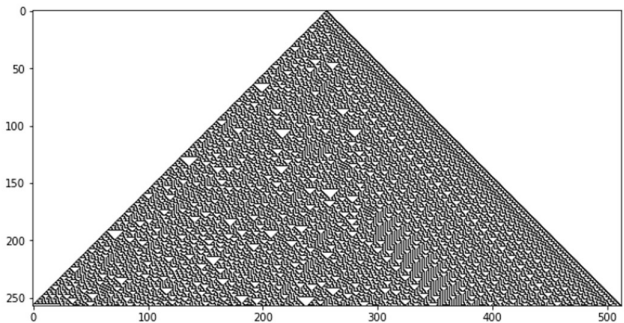

Fig 5. Reverse Rule 30 for 256 iterations

Thus, we can clearly see how decisive branching is giving rise to randomness. So, we have seen that a slight modification of Equation (ii) is giving rise to randomness. An interesting observation is that, we can subside the decisive branching in Rule 30 by suppressing Equation (iii). This can be done by ANDing Equation (i) and (ii) ; that is we are ANDing Rule 30 with its symmetric counter-part. The net output is very interesting and it is given in Figure 6. It is the figure of Rule 90 which is again symmetric and systematic.

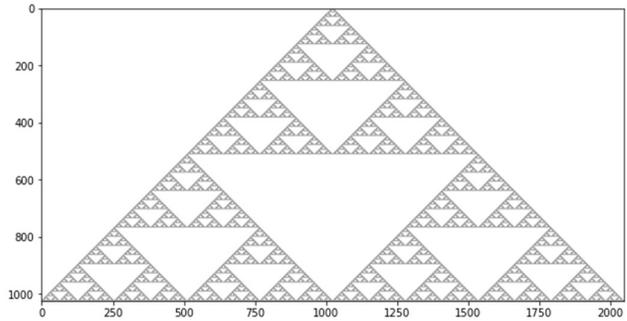

Fig 6. Rule 90 generated by ANDing Rule 30 with Rule 150

Thus, by various arguments we have proved that Decisive Branching is the cause of randomness in Elementary Cellular Automaton, specifically Rule 30.

3. DOES THIS RANDOMNESS SUBSIDES

Now we have understood how decisive branching is causing the randomness in Rule 30. We will now find out the nature of the randomness. We have defined a parameter called randomness count which is the ratio of how many times the right cell and center cell both are equal to one; to the total number of Boolean operations which needs to be done for an evaluation of Rule 30.

*Randomness Count*

$$= \frac{number\ of\ boolean\ calculation\ where\ right\ cell\ =\ centre\ cell\ =\ 1}{total\ number\ of\ boolean\ calculation}$$

------- (vi)

We will plot this randomness count for 1024 iterations and find out the nature of the curve. Fig 7 shows the randomness count versus the number of bits or the length of the Rule 30 vector. For the last $1024^{th}$ iteration, the number of Boolean calculation where right and center cell equal one came out to be 261648 and the total number of Boolean calculations comes out to be 1836528. This gives a randomness count of approximately 0.14245. Analyzing fig 7 we can see the randomness count has some large variations (overshoot in the plot) for small number of iterations but for large iteration, it is becoming stable and tries to reach a steady state value which is in the vicinity of 0.14

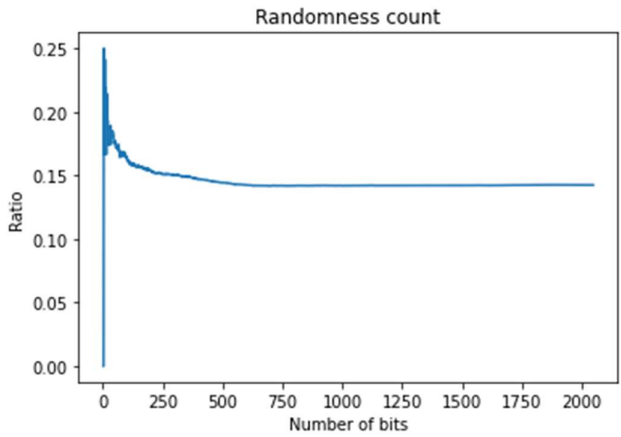

Fig 7. randomness count versus the number of bits.

### 4. Relation between randomness and number of ones and zeroes

The second question of Wolfram Rule 30 challenge addresses whether the ratio of total number of one and zero approaches exactly one or not. This ratio plays an important role in the randomness as the nature of this ratio has correlation with the Randomness Count Graph shown in Fig 7. Fig 8 shows the variation of the ratio of number of zeroes to ones till 2048 bits or 1024 iterations.

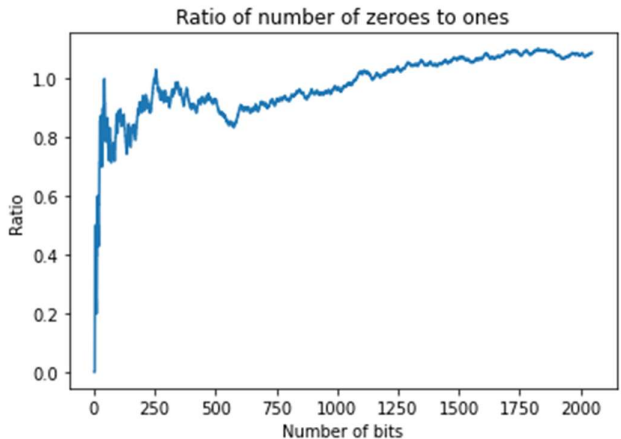

Fig 8. Ratio of zeroes to ones

Fig 9 shows the two graphs: Ratio of zero and one, with the randomness count graph. The Randomness Count Graph has been scaled up for comparison and correlation purposes. We can see that the two graphs in Fig 9 have a moderate correlation between them. Pearson Correlation Coefficient for the two of them comes out to be 0.45. There is an upshoot in both of the curves for small iterations, whereas for large values of iteration both of them try to reach a steady state value.

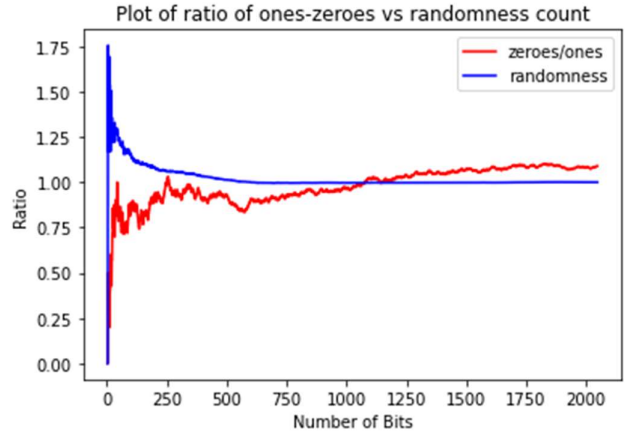

Fig 9. Ratio of ones-zeroes vs randomness count

### 5. Conclusion to problem 1

Now that we have seen correlation between the two plots, we can say that the ratio of zeroes/ones and randomness count are mildly correlated. In this particular case, correlation equals causation, simply because the degree of randomness will affect the number of zero and one. Now the ratio of zeroes and ones approaches one which is shown in first one billion iteration in the Wolfram dataset. We can make this statement that the ratio of zeroes and ones can never approach zero in the steady state. Now as the Randomness Count is correlated with that of the ratio of zero/one. We can state that the Randomness count can never be zero in steady state. By simulation we have seen that it actually approaches a steady state value of 0.14. Table 2 shows the steady state values for various iterations. Now since the randomness count can never be zero, the central value of the Rule 30 structure will always remain random as there will always be some decisive branching taking place. Thus, as the central column continues to remain purely random it will always be aperiodic.

| Number of iterations | Randomness Count |
|---|---|
| 1024 | 0.1424688324926165 |
| 2048 | 0.1429953891438367 |
| 4096 | 0.14306875512717437 |
| 8192 | 0.14293569508508577 |
| 16384 | 0.14282179825815405 |

Table 2: Steady State Values for randomness count